 \documentclass[final,5p,times,twocolumn]{elsarticle}

\usepackage{amssymb}
\usepackage{multirow}
\usepackage{amsmath}
\usepackage{doi}
\usepackage{enumitem}
\usepackage{listings,multicol}
\usepackage{parcolumns}
\usepackage{xcolor}

\definecolor{codegreen}{rgb}{0,0.6,0}
\definecolor{codegray}{rgb}{0.5,0.5,0.5}
\definecolor{codepurple}{rgb}{0.58,0,0.82}
\definecolor{backcolour}{rgb}{0.95,0.95,0.92}

\lstdefinestyle{mystyle}{
    backgroundcolor=\color{backcolour},   
    commentstyle=\color{codegreen},
    keywordstyle=\color{magenta},
    numberstyle=\tiny\color{codegray},
    stringstyle=\color{codepurple},
    basicstyle=\ttfamily\footnotesize,
    breakatwhitespace=false,         
    breaklines=true,                 
    captionpos=b,                    
    keepspaces=true,                 
    numbers=left,                    
    numbersep=5pt,                  
    showspaces=false,                
    showstringspaces=false,
    showtabs=false,                  
    tabsize=2
}

\usepackage{booktabs}
\usepackage{caption} 
\usepackage{graphicx,subfigure}
\usepackage{verbatim}
\usepackage{makecell}
\usepackage{longtable}

\usepackage{rotating}

 \usepackage{colortbl} 
 \usepackage[htt]{hyphenat}

\usepackage{lineno}

\begin{document}

\begin{frontmatter}

\title {SSLG4: A Novel Scintillator Simulation Library for Geant4}


\author[label1]{Mustafa Kandemir\corref{cor1}}
\cortext[cor1]{corresponding authors}
\ead{mustafa.kandemir@erdogan.edu.tr}
\author[label2,label3]{Emrah Tiras\corref{cor1}}
\ead{etiras@fnal.gov}
\author[label2]{Burcu Kirezli}
\author[label4]{İbrahim Koca}

\address[label1]{Department of Physics, Recep Tayyip Erdogan University, Rize, 53100, Türkiye}
\address[label2]{Department of Physics, Erciyes University, Kayseri, 38030, Türkiye}
\address[label3]{Department of Physics and Astronomy, The University of Iowa, Iowa City, IA, 52242, USA}
\address[label4]{Department of Astronomy and Space Sciences, Erciyes University, Kayseri, 38030, Türkiye}




\begin{abstract} 

This study introduces a new Scintillator Simulation Library called SSLG4 for the Geant4 Monte Carlo simulation package. With SSLG4, we aim to enhance efficiency and accelerate progress in optical simulations within the Geant4 framework by simplifying scintillator handling and providing a rich repository of scintillators. The SSLG4 enables users to quickly include predefined scintillator materials in their simulations without requiring manual definition. The library initially contains 68 scintillators, consisting of 58 organic and 10 inorganic types. Most of these scintillators are selected from the catalogs of several scintillator manufacturers, notably Eljen and Luxium. Other scintillators are included based on their widespread use across various physics domains. The library stores optical data of scintillators in ASCII files with .mac and .txt extensions, enabling users to add, remove, or modify properties of scintillators at runtime of their applications. In addition, we made all the scintillator data available in the library on a dedicated page of our website to ensure convenient access for all users. 

\textbf{Program summary}  \\
\textit{Program title:} SSLG4 \\
\textit{CPC Library link to program files:} \\
\textit { Developer's repository link:} \url{https://github.com/mkandemirr/SSLG4} \\
\textit {Website link:} \url{https://neutrino.erciyes.edu.tr/SSLG4/} \\
\textit{Licensing provisions:} GNU General Public License 3 \\
\textit{Programming language:} C++ \\
\textit{Operating system:} Cross-Platform \\
\textit{External routines/libraries:} Geant4, CMake, OPSim \\
\textit{Nature of problem:} \\
Defining a new scintillator in Geant4 is a cumbersome process for some users due to three main reasons: (1) It requires a lot of data input from users, (2) collecting the scintillator data requires an extensive literature review, and (3) the collected data needs to be converted into the desired format. In addition, the interfaces provided to define a scintillator direct users to embed scintillator data into their source code, resulting in increased code complexity, reduced code readability, and an inefficient working environment.
\textit{Solution method:} \\
To solve the problems mentioned above, developing and introducing a new library consisting of fully parameterized and ready-to-use scintillators would greatly increase the useability of the Geant4 simulation package for scintillator studies and interest a wide range of scientific communities.

\end{abstract}


\end{frontmatter}


\section{Introduction}
\label{sec1}

Scintillation detectors play a pivotal role in particle and nuclear physics due to their remarkable ability to detect and quantify ionizing radiation. By converting the energy deposited by particles into detectable flashes of light, they provide precise measurements of particle properties such as energy, momentum, and type. This information is crucial in identifying new particles, studying their characteristics, and testing theories such as the Standard Model of particle physics and beyond. In addition, due to their versatility, adaptability, and efficiency in detecting ionizing radiation, their application domains span numerous fields including medical imaging, nuclear medicine, homeland security and defense, environmental monitoring, and space exploration \cite{Lecoq}.

Monte Carlo simulation of optical photons is an indispensable part of accurately modeling the behavior of scintillation detectors. Among the tools with limited optical photon transport capability, Geant4 \citep{Agostinelli,G4Toolkit, G4App} is the most widely used one for this purpose due to its detailed treatment of optical photon interactions with both materials and surfaces, flexibility in simulation control, its open-source nature, and extensive documentation.

In our previous study, we addressed the challenges encountered by many users when developing optical applications within Geant4. To overcome these challenges, we introduced OPSim \cite{opsim}, a set of additions to the Geant4 toolkit. OPSim includes several C++ classes that make it easy to define optical components for scintillation detectors. Moreover, OPSim provides abstract interface classes designed to encourage users to develop reusable and portable material build codes, which are independent of their applications while working on their projects. In this study, we created a scintillator library, which we named Scintillator Simulation Library for Geant4 (SSLG4), utilizing these interfaces. 

SSLG4 comprises fully parametrized ready-to-use scintillators. The advantages of using this library are detailed below:

\begin{itemize}
\setlength\itemsep{0.em}

\item Obtaining a scintillator with SSLG4 is remarkably simple and requires only a single line of code. This process is similar to obtaining a predefined material from the NIST database in Geant4.

\item As the scintillators in SSLG4 are fully parameterized, users have complete control over all optical properties at runtime of their applications.

\item SSLG4 improves the clarity and maintainability of user detector construction code by decoupling the data necessary for defining scintillators from the source code. This approach aligns with the SOLID principles of Object-oriented Design \cite{bobUncle}.

\item SSLG4 helps to create a flexible and efficient simulation environment, particularly for applications involving multiple scintillators.

\item Since SSLG4 stores scintillator data in a format directly accepted by Geant4, users who do not want to use the SSLG4 software in their applications can also benefit from this data.

\item SSLG4 can be a valuable reference source as it collects a wide range of scintillator data in one place. To illustrate, in many studies involving scintillator simulations, published simulation results depend heavily on the input parameters used. However, specifying the parameters used in the simulation one by one can be a bit complicated. In this regard, SSLG4 can serve as a resource where these input parameters can be easily cited. Such a resource also allows accurate comparison of simulation results from different studies.

In Section 2, we will delve into the implementation details of SSLG4. Following this, in Section 3 we will give an overview of SSLG4 and outline future goals. Finally, in Section 4, we will present an example Geant4 application developed to demonstrate the correct transfer of scintillator data from SSLG4 to the Geant4 system.

\end{itemize}

\section{Implementation}

In this research, we first extended OPSim by adding a new class called \texttt{ScintillatorBuilder}, which allows building scintillators using various constructors. We then extensively utilize this class to create scintillators within SSLG4. Listing~\ref{lst1} shows two distinct ways of building a scintillator with OPSim.

\begin{figure*}
\begin{lstlisting}[label=lst1, language={C++}, caption={Two different ways of creating a scintillator using OPSim.}]
int main()
{
  G4Material *scntMat = new G4Material(...);
  const G4String macroFilePath = "path/to/macro/file.mac";
  //Two different ways to create a scintillator using OPSim.
  {//First method                   
    MaterialPropertiesTable* mpt = new MaterialPropertiesTable("ej-301");
    G4UImanager::GetUIpointer()->ExecuteMacroFile(macroFilePath);
    scntMat->SetMaterialPropertiesTable(mpt);
  } 
  {//Second method. Available in the new version of OPSim.
    ScintillatorBuilder builder("ej-301", macroFilePath, scntMat);
    scntMat = builder.GetProduct();
  }
}


\end{lstlisting}
\end{figure*}

\begin{figure*}[!htp]
 \centering
    \includegraphics[width=0.7\textwidth]{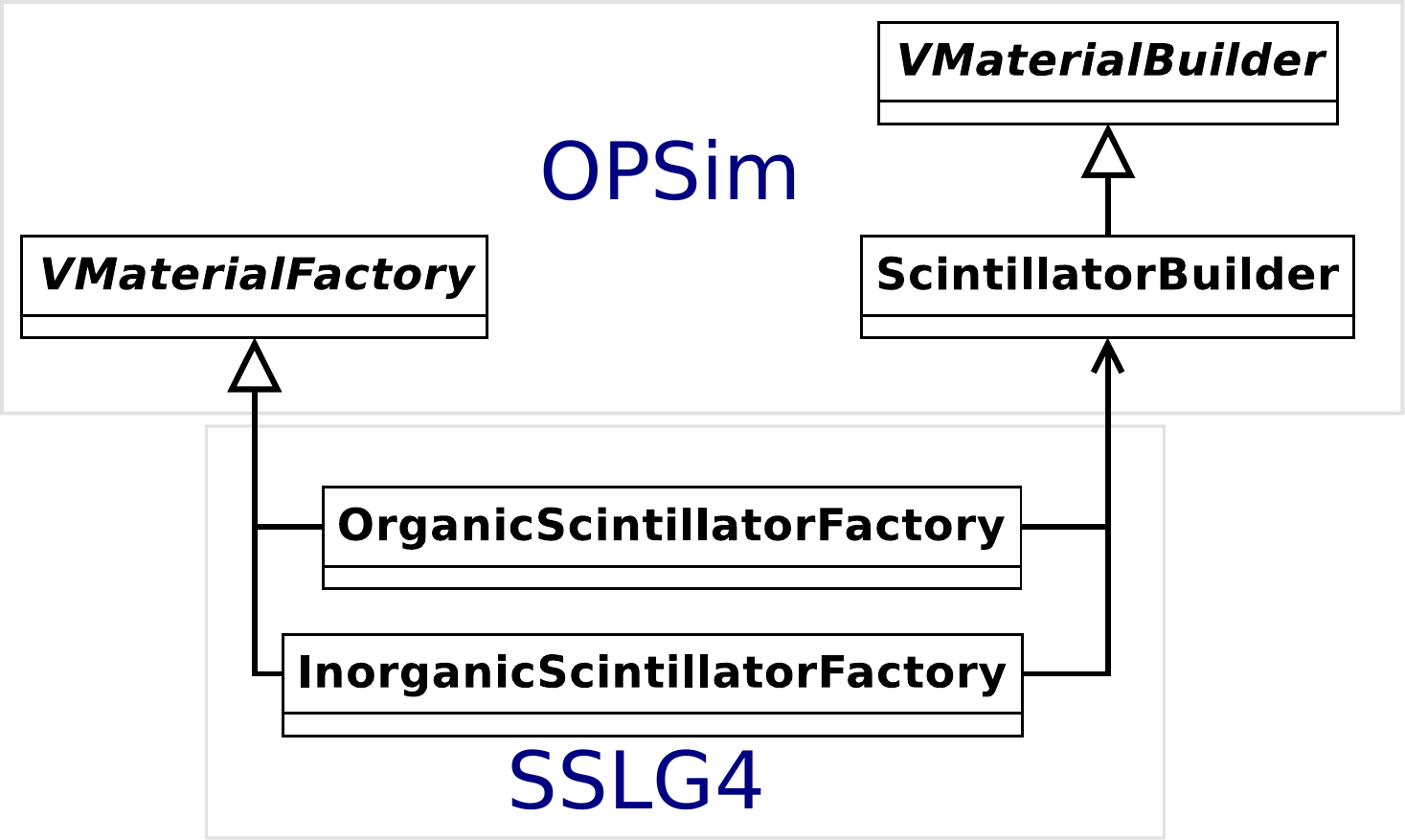}
     \caption{ UML \cite{Fowler:2015} diagram of SSLG4 and its relation with OPSim \cite{opsim}. Only the relevant classes of OPSim are shown in the figure.  }
    \label{fig1}
\end{figure*}

SSLG4 is developed by inheriting from the \texttt{VMaterialFactory} class provided by OPSim. It uses all the functionalities of OPSim and is therefore heavily dependent on it. Figure \ref{fig1} illustrates the structure of SSLG4 with a UML \cite{Fowler:2015} diagram. As depicted in the figure, there is a scintillator factory class for both organic and inorganic scintillator types. 

Users need to utilize one of these two singleton factory classes to obtain a scintillator. The example provided in Listing~\ref{lst2} illustrates how to use these classes on the user side.

\begin{figure*}
\begin{lstlisting}[label=lst2, language={C++}, caption={Getting a predefined scintillator from SSLG4. Users can change the properties of the scintillators at runtime of their application.}]
int main()
{
  G4bool isMPTOn = true;//MPT = MaterialPropertiesTable
  //Getting an organic scintillator object 
  G4String osCode = "OPSC-100"; 
  G4Material* scntMat1 = OrganicScintillatorFactory::GetInstance()->Get(osCode, isMPTOn);
  //Getting an inorganic scintillator object
  G4String isCode = "ISC-1000";
  G4Material* scntMat2 = InorganicScintillatorFactory::GetInstance()->Get(isCode, isMPTOn);
}
\end{lstlisting}
\end{figure*}

SSLG4 is designed to be flexible, so adding a new scintillator is fairly straightforward and can be done in two different ways. The first method is to use one of the appropriate constructors of the \texttt{ScintillatorBuilder} class within the factory class to which the scintillator belongs. This can be easily accomplished by examining the implementation of other scintillators in that class. The second method is to inherit directly from the \texttt{VMaterialBuilder} class in OPSim. In this approach, only the object of the scintillator is created within the factory class. This method is particularly suitable for scintillators with complex chemical compositions or implementations that necessitate lengthy code.

\section{Overview of SSLG4}
\label{sec2}

SSLG4 currently contains 68 scintillators, comprising 58 organic and 10 inorganic types. Table 1 provides a complete list of the scintillators available in the library. Each scintillator in the library is assigned a unique code, as shown in the last column of the table. Users should utilize these codes to access their desired scintillators in their Geant4 applications.

Many of the scintillators in the library have been selected from the catalogs of several scintillator manufacturers, particularly Eljen and Luxium. Other scintillators are included due to their widespread use across various physics domains.

The scintillator data in the library have been collected from diverse sources. For commercially available scintillators, the data are extracted from published data sheets provided by the respective manufacturers, with a few exceptions. For domain-specific scintillators, we obtain data from GitHub repositories maintained by competent individuals or well-known research groups within those fields.

Before collecting the scintillator data, we surveyed Geant4's official forum site \cite{G4forum} to determine which scintillator properties users most required for their simulations. Additionally, we examined numerous Geant4 applications and projects on GitHub repositories, encompassing both large-scale and smaller-scale projects. Our findings indicated that, although Geant4 offers a wide array of properties for scintillators, users primarily needed the following properties, which varied depending on the physics of interest: SCINTILLATIONCOMPONENT1, SCINTILLATIONYIELD, RINDEX, ABSLENGTH, and SCINTILLATIONTIMECONSTANT1. We took particular care to ensure that these properties were included while collecting scintillator data.

In addition, to ensure a wider community can benefit from the data in the library, we have published all scintillator data on a dedicated page of our website in a format accepted by Geant4. 

Our long-term goal is to increase the number of scintillators in the library and make more scintillator property data available to users.

\clearpage
\onecolumn

\begin{longtable}{llll}
\caption{Complete list of scintillators available in SSLG4. Each scintillator in the library has a unique code. This is shown in the last column of the table. Users should use these codes to include the desired scintillators in their Geant4 applications.}
\label{tab:my-table}\\
\hline
\textbf{Type} & \textbf{Identification} & \textbf{Name} & \textbf{Simulation Code (SC)} \\ \hline
\endfirsthead
\multicolumn{4}{c}%
{{\bfseries Table \thetable\ continued from previous page}} \\
\endhead
%
\endfoot
\endlastfoot
\textbf{Organic Plastic} & \textbf{Eljen Technology \cite{Eljen}/Luxium Solutions \cite{Luxium}} & EJ-200/Pilot F/BC-408 & OPSC-100 \\
 &  & EJ-204/NE -104/BC-404 & OPSC-101 \\
 &  & EJ-208/NE -110 /BC-412 & OPSC-102 \\
 &  & EJ-212/NE-102A/BC-400 & OPSC-103 \\
 &  & EJ-214 & OPSC-104 \\
 &  & EJ-228/Pilot U/BC-418 & OPSC-105 \\
 &  & EJ-230/Pilot U2/BC-420 & OPSC-106 \\
 &  & EJ-232/NE-111A/BC-422 & OPSC-107 \\
 &  & EJ-232Q-0.5 & OPSC-108 \\
 &  & EJ-232Q-1.0 & OPSC-109 \\
 &  & EJ-232Q-2.0 & OPSC-110 \\
 &  & EJ-232Q-3.0 & OPSC-111 \\
 &  & EJ-232Q-5.0 & OPSC-112 \\
 &  & EJ-240/NE-115/BC-444 & OPSC-113 \\
 &  & EJ-244/BC-440 & OPSC-114 \\
 &  & EJ-244M/BC-440M & OPSC-115 \\
 &  & EJ-248/BC-448 & OPSC-116 \\
 &  & EJ-248M & OPSC-117 \\
 &  & EJ-254-1pct & OPSC-118 \\
 &  & EJ-254-2.5pct & OPSC-119 \\
 &  & EJ-254-5pct & OPSC-120 \\
 &  & EJ-256-1.5pct & OPSC-121 \\
 &  & EJ-256-5pct & OPSC-122 \\
 &  & EJ-260/NE-103/BC-428 & OPSC-123 \\
 &  & EJ-262 & OPSC-124 \\
 &  & EJ-276D & OPSC-125 \\
 &  & EJ-276G & OPSC-126 \\
 &  & EJ-280 & OPSC-127 \\
 &  & EJ-282 & OPSC-128 \\
 &  & EJ-284 & OPSC-129 \\
 &  & EJ-286 & OPSC-130 \\
 &  & EJ-290/BC-490/NE-120 & OPSC-131 \\
 &  & EJ-296/BC-498 & OPSC-132 \\
 &  & EJ-426 & OPSC-133 \\ \cline{2-4} 
 & \textbf{Nuviatech Instruments \cite{Nuviatech} } & SP-32 & OPSC- 200 \\
 &  & SP-33 & OPSC- 201 \\ \cline{2-4} 
 & \textbf{Hangzhou Shalom EO \cite{Shalom} } & SP-102 & OPSC- 300 \\ \cline{2-4} 
 & \textbf{Rexon Components \cite{Rexon} } & RP-408 & OPSC-400 \\ \hline
\textbf{Organic Liquid} & \textbf{Eljen Technology/Luxium Solutions} & EJ-301/NE-213/BC-501A & OLSC-100 \\
 &  & EJ-309 & OLSC-101 \\
 &  & EJ-309B-1pct & OLSC-102 \\
 &  & EJ-309B-2.5pct & OLSC-103 \\
 &  & EJ-309B-5pct & OLSC-104 \\
 &  & EJ-313/NE-226/BC-509 & OLSC-105 \\
 &  & EJ-315-H/BC-537/NE-230 & OLSC-106 \\
 &  & EJ-321H & OLSC-107 \\
 &  & EJ-321L & OLSC-108 \\
 &  & EJ-321P & OLSC-109 \\
 &  & EJ-321S & OLSC-110 \\
 &  & EJ-325A & OLSC-111 \\
 &  & EJ-331-0.5pct/NE-323/BC-521 & OLSC-112 \\
 &  & EJ-335-0.25pct/BC-525 & OLSC-113 \\
 &  & EJ-351/NE-220/BC-573 & OLSC-114 \\ \cline{2-4} 
 & \textbf{HEP Materials (Neutrino Studies) \cite{Rat} } & WbLS-1pct & OLSC- 200 \\
 &  & WbLS-1pct-gd-0.1pct & OLSC- 201 \\
 &  & WbLS- 3pct & OLSC- 202 \\
 &  & WbLS- 3pct-gd-0.1pct & OLSC- 203 \\
 &  & WbLS-5pct & OLSC- 204 \\ \hline
\textbf{Inorganic} & \textbf{Advatech \cite{Advatech} } & BaF2 & ISC-1000 \\
 &  & CdWO4 & ISC-1001 \\ \cline{2-4} 
 & \textbf{Luxium Solutions} & BGO & ISC- 2000 \\
 &  & CsINa & ISC- 2001 \\
 &  & CsITI & ISC- 2002 \\
 &  & LYSOCe & ISC- 2003 \\
 &  & NaITI & ISC- 2004 \\ \cline{2-4} 
 & \textbf{HEP Materials (Noble gases) \citep{Hans, Underground, Sarah} } & LAr & ISC- 3000 \\
 &  & LXe & ISC- 3001 \\
 &  & PbWO4 & ISC- 3002 \\ \hline
\end{longtable}

\clearpage
\twocolumn

\section{Testing the framework of SSLG4}

In this section, we present a Geant4 application developed to test whether the properties of scintillators defined in SSLG4 are accurately transferred to the Geant4 system. The application is entirely controlled by a macro file, which contains various simulation settings. These settings include scintillator selection, primary particle control, toggling optical physics processes on and off, controlling critical optical parameters, and specifying the simulation output file type. The application outputs the following physics quantities in an n-tuple format on an event-by-event basis:

\begin{itemize}
\setlength\itemsep{0.em}
\item Event ID
\item Wavelength and energy spectrum of emitted photons.  
\item Photon emission time spectrum. 
\item Number of emitted scintillation photons. 
\item Number of emitted cherenkov photons. 
\item Total energy deposited. This parameter is included for confirmation, as the number of photons emitted depends on the deposited energy.
\end{itemize}

As these quantities are retrieved from the Geant4 kernel during the simulation stage and are generated based on the scintillator data provided by SSLG4, they can be used, in a sense, to check the proper functioning of the SSLG4 system.

\begin{figure*}[!htp]
 \centering
    \includegraphics[width=0.9\textwidth]{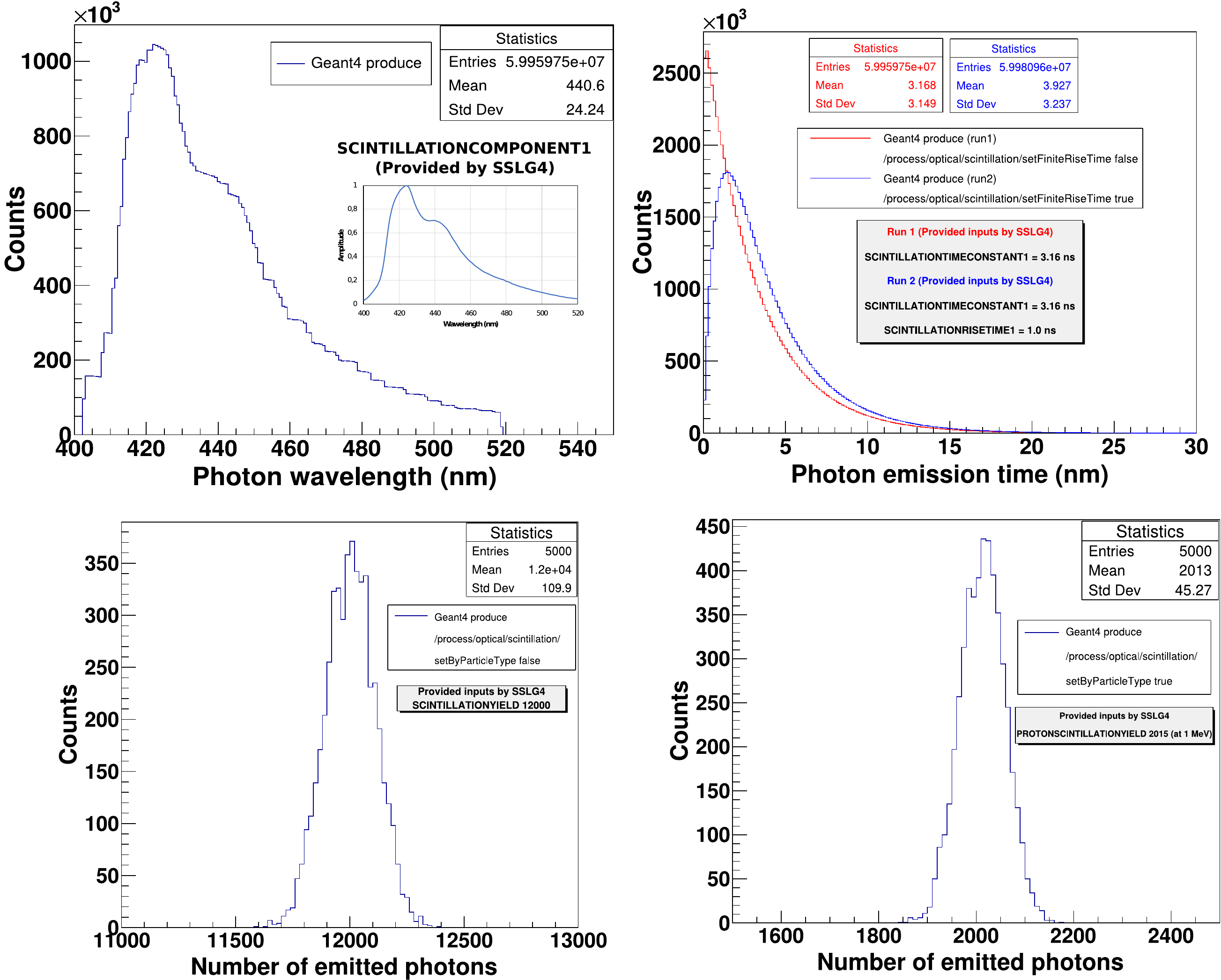}
     \caption{ The top two graphs and the bottom left graph are obtained from the 1 MeV electron simulation, while the bottom right one is obtained from the 1 MeV proton simulation. These graphs illustrate several properties defined for EJ-301 in SSLG4 and what the Geant4 system produces based on these provided properties.  }
    \label{fig2}
\end{figure*}

We selected EJ-301 from SSLG4 as an example to demonstrate the accurate transfer of its properties to the Geant4 framework. Since there are numerous properties for a scintillator, we chose only a few, considering the properties most needed in a typical optical application. These include SCINTILLATIONCOMPONENT1, SCINTILLATIONYIELD, SCINTILLATIONTIMECONSTANT1, and SCINTILLATIONRISETIME1. These properties are independent of particle type. Additionally, we choose a particle-dependent property, PROTONSCINTILLATIONYIELD, for scintillator simulation based on particle type. Figure \ref{fig2} summarizes the simulation results obtained from the 1 MeV electron and 1 MeV proton simulations. 

\section{Conclusions}

This paper presents SSLG4, a novel library designed to simplify scintillator usage within Geant4 for optical applications. SSLG4 seeks to improve efficiency and speed up progress in optical simulations within the Geant4 framework by simplifying scintillator handling and offering a large repository of scintillators. With SSLG4, users can effortlessly acquire a scintillator as a \texttt{G4Material} object using just one line of code, akin to obtaining predefined materials from the \texttt{G4NistManager} class. 

The library encompasses a diverse range of scintillators, including both organic and inorganic types, commonly utilized in high-energy and nuclear physics experiments. Initially, the library comprises 68 scintillators, including 58 organic and 10 inorganic types. Most of these scintillators are sourced from the catalogs of prominent scintillator manufacturers such as Eljen and Luxium. Other scintillators are included based on their widespread usage across various physics domains.

SSLG4 stores optical data for each scintillator in ASCII files with extensions .mac and .txt. This design allows users to dynamically modify scintillator properties during runtime by adding, removing, or adjusting parameters. 

Additionally, we published detailed scintillator data in a format accepted by Geant4 on a dedicated page of our website, ensuring easy access for researchers and developers.

\section{Acknowledgement}

This study was funded by the Scientific and Technological Research Council of Türkiye (TÜBİTAK) ARDEB 1001 Grant Number 122F095, and supported by the Scientific Research Projects (BAP) of Erciyes University, Türkiye, under the grant contracts of FBAÜ-2023-12325, FBG-2022-11499, and FDS-2021-11525. Dr. Emrah Tiras is thankful for the support of the Turkish Academy of Sciences (TUBA) under the Outstanding Young Scientists Awards Program (GEBIP) grant. The authors would like to thank the Office of the Dean for Research for providing the Lab’s infrastructure at the ARGEPARK building of Erciyes University and the Proofreading \& Editing Office for proofreading this manuscript.




\clearpage



\bibliographystyle{elsarticle-num}

\bibliography{sample}

\end{document}